\documentclass[showkeys,aps,pre,longbibliography,reprint,superscriptaddress,floatfix,notitlepage]{revtex4-2}

\usepackage[dvipsnames]{xcolor}
\definecolor{linkcolor}{rgb}{0.3,0.3,1.0} %hyperlink
\usepackage[pdftex,colorlinks=true, linkcolor= linkcolor, citecolor= linkcolor, urlcolor= linkcolor, hyperindex=true,hyperfigures=true]{hyperref} %hyperlink
\usepackage{amsmath,amssymb}
\usepackage{graphicx}
\usepackage{color}
\usepackage{subfigure}

\begin{document}

%%%
\title{Physical properties of a generalized model of multilayer adsorption of dimers}
%\title{{\color{green}Scaling relations in a generalized model of multilayer adsorption of dimers}}
%%%

%%%
\author{G Palacios}
\affiliation{CRCN-NE/CNEN, 50740-545, Recife, PE, Brazil}
\email{palaciosg226@gmail.com}

\author{Sumanta Kundu}
\affiliation{Department of Physics and Astronomy, University of Padova, 
Via Marzolo 8, I-35131 Padova, Italy}
\affiliation{INFN, Sezione di Padova, Via Marzolo 8, I-35131 Padova, Italy}

\author{L A P Santos}
\affiliation{CRCN-NE/CNEN, 50740-545, Recife, PE, Brazil}
\affiliation{SCIENTS, 53635-015, Igarassu, PE, Brazil}

\author{M A F Gomes}
\affiliation{Departamento de Física, Universidade Federal de Pernambuco, 50670-901, Recife, PE, Brazil}

\begin{abstract}
%This article investigates the theoretical physics aspects and, in particular, the statistical properties of two-dimensional structures obtained with a dimer packing model, which additionally may be of great interest in other scientific and technological areas. The study is based on very extensive computer simulations in order to examine the effect of the orientational anisotropy of dimers deposited following a ballistic deposition process (aspects not yet examined in these deposition models), leading to multilayer networks of complex morphologies dominated by branched fractal architectures. The geometric characteristics of the bulk and contours of these systems, and the transport properties, particularly the electrical conductivity were investigated by varying the orientational anisotropy of the dimers. The results provide information about the fundamental mechanisms underlying formation and behavior of such types of amorphous and disordered matter that are of paramount importance both to basic physics as well as to environmental and material sciences.

We investigate the transport properties of a complex porous structure with branched fractal architectures formed due to the gradual deposition of dimers in a model of multilayer adsorption. We thoroughly study the interplay between the orientational anisotropy parameter $p_0$ of deposited dimers and the formation of porous structures, as well as its impact on the conductivity of the system, through extensive numerical simulations. By systematically varying the value of $p_0$, several critical and off-critical scaling relations characterizing the behavior of the system are examined. The results demonstrate that the degree of orientational anisotropy of dimers plays a significant role in determining the structural and physical characteristics of the system. We find that the Einstein relation relating to the size scaling of the electrical conductance holds true only in the limiting case of $p_0 \to 1$. Monitoring the fractal dimension of the interface of the multilayer formation for various $p_0$ values, we reveal that in a wide range of $p_0 > 0.2$ interface shows the characteristic of a self-avoiding random walk, compared to the limiting case of $p_0 \to 0$ where it is characterized by the fractal dimension of the backbone of ordinary percolation cluster at criticality. Our results thus can provide useful information about the fundamental mechanisms underlying the formation and behavior of wide varieties of amorphous and disordered systems that are of paramount importance both in science and technology as well as in environmental studies.

\end{abstract}

\maketitle
\newpage

%\keywords{random sequential adsorption, multilayer, percolation, conductivity}

\section{\label{sec:level1}Introduction}
Understanding the transport properties of charges/fluids in complex disordered/porous media is of paramount interest in different disciplines of science as it provides deep insights into many natural and industrial processes involving the flow of particles/fluids through a network of connected units or pores. Relevant examples include fluid flow in sedimentary rocks, transport of contaminants in underground water, radioactive nuclei transport from nuclear waste disposal repositories, and transport of colloidal particles in vascular systems~\cite{Woods,Dietrich2005,Sahimi-book2011,Xiong,Bourg2003,Bourg2006,Bassingthwaighte}. In these systems, the conductivity of the largest connected component of the porous network determines the transport properties.

Importantly, the conductivity of the system is known to be dependent on the micro-structural properties of the porous network~\cite{Quenard,Wei,printsypar,Wang2020}, while the structure itself gets modified due to the sedimentation or precipitation of particles transported through fluids~\cite{Sahimi-book2011,Feder2022}. This often gives rise to phenomena like clogging~\cite{Schwartz1993,Herrmann2018,Agbangla2014}. At the pore scales, the occurrence of such a process might induce dysfunctionalities in many biological processes and can be dangerous, contrarily, it is utilized for the purification of gases and liquids and for the separation of various important compounds.

In many natural systems, the pore network structure appears to be statistically self-similar over several length scales, such as porous rocks or sedimentary reservoirs, and vascular systems, whose complex internal structures are characterized by fractal geometries~\cite{Bassingthwaighte,Wei,FERANIE}. The knowledge about the transport properties in these systems is advantageous for many practical applications, for example, oil recovery from geological media and storage of gases such as hydrogen or activated carbons~\cite{Blankenship,Dincer,Heinemann,Morris}.

The percolation theory has long been served as the starting point to study the transport properties in all these systems~\cite{broadbent1957,Stauffer2018,Sahimi-book2023}. The emergence of a macroscopically large connected component at the percolation threshold and its fractal nature has provided basic information about the phase transition and critical behavior of the system. The conductivity of the percolation backbone has also been investigated by introducing different models of percolation and taking into account different types of heterogeneity in the system~\cite{Sampaio2018,Deng_2022}.

A central question that is still not fully understood is whether and to what extent the micro-structural details of fractal porous media formed due to the gradual deposition of particles (e.g., sedimentary rock formation) affect the transport properties of the system, and whether they are universal in nature. Motivated by this, we consider one of the variants of the ballistic deposition model~\cite{Meakin1986,Meakin1990,Yu2002,comets2022} that produces a growing structure with variability in pore-size distribution and study in detail the conductivity properties of the system by means of numerical simulations.

In the simplest case of the ballistic deposition model~\cite{Meakin1986,Meakin1990}, particles in the form of monomers are dropped one by one onto an initially empty flat substrate at random and stick irreversibly to the surface of the growing structure. A newly released particle can sit only on top of an already existing particle in the presence of excluded volume interactions. Additionally, with a non-zero sticking probability to the side of the nearest neighbor particle, the model is capable of generating porous structures. In general, the dynamics leads to the formation of a particle aggregate with complex structural properties. The roughness of the growing interface or the active zone exhibits nontrivial properties described by the Family-Vicsek scaling~\cite{Family_1985,Kardar1986}. Furthermore, it has been shown that the active zone is not a self-similar fractal~\cite{Meakin1986}.

In this paper, we generalize our recently introduced ballistic deposition model of extended objects for multilayer adsorption \cite{Palacios2022} to generate complex internal porous structures that can be controlled by varying the parameters of the model. Here, the dimers (horizontal/vertical) are dropped vertically and are deposited on the top of a randomly selected column with growing structure formation and thus, in turn, may prevent from accessing the lower layers at the selected column for future deposits, generating pores in the structure. By systematically controlling the orientational anisotropy $p_0$ of the dimers, we thoroughly investigate the conductivity of the system. Moreover, we study the system from the perspective of critical phenomena and report scaling relations of several physical observables in the vicinity of the critical point. Interestingly, the Einstein relation associated with the critical exponents characterizing the fractal structure seems to hold only for $p_0$ close to unity for the system sizes considered here.

The structure of the paper is organized as follows: In Section \ref{sect:Details}, we give the details of our simulation; in Section \ref{sect:Results}, the main results are presented and a discussion is made on the morphological details of the structure, including the percolation transition (\ref{sect:Percolation}), the fractality and diffusion on the percolation clusters (\ref{sect:Fractal}), the electrical conductivity (\ref{sect:cunct}) and the fractal properties of the interface (\ref{sect:Interface}). In Section \ref{sec:Canclu}, we summarize our findings and present some future perspectives.

\section{Model and simulation details} \label{sect:Details}

We consider a variant of the ballistic deposition model for the formation of a growing structure on a one-dimensional lattice of size $L$ with hard wall boundary conditions. Dimers are dropped one by one onto the randomly selected lattice sites from a far distance from the growing structure along the vertical direction. Specifically, a dimer follows a vertical trajectory from its release point and lands on the top of the growing surface where it first encounters a previously deposited dimer. A dimer occupies two consecutive lattice sites. For simplicity, we neglect the diffusive motion of the incoming dimers. Note that the dimers are non-sticky in nature and therefore, during its vertical motion even if a dimer finds a previously deposited dimer at the side of the nearest neighbor columns, it continues its motion until it hits a dimer beneath it and it cannot go down. The dynamics leads to a complex porous structure formation due to the orientational anisotropy of the incoming dimers.

\begin{figure}[t]
    \centering  
    %\includesvg[width=0.72\columnwidth]{sketch1.svg} 
    \includegraphics[width=0.72\columnwidth]{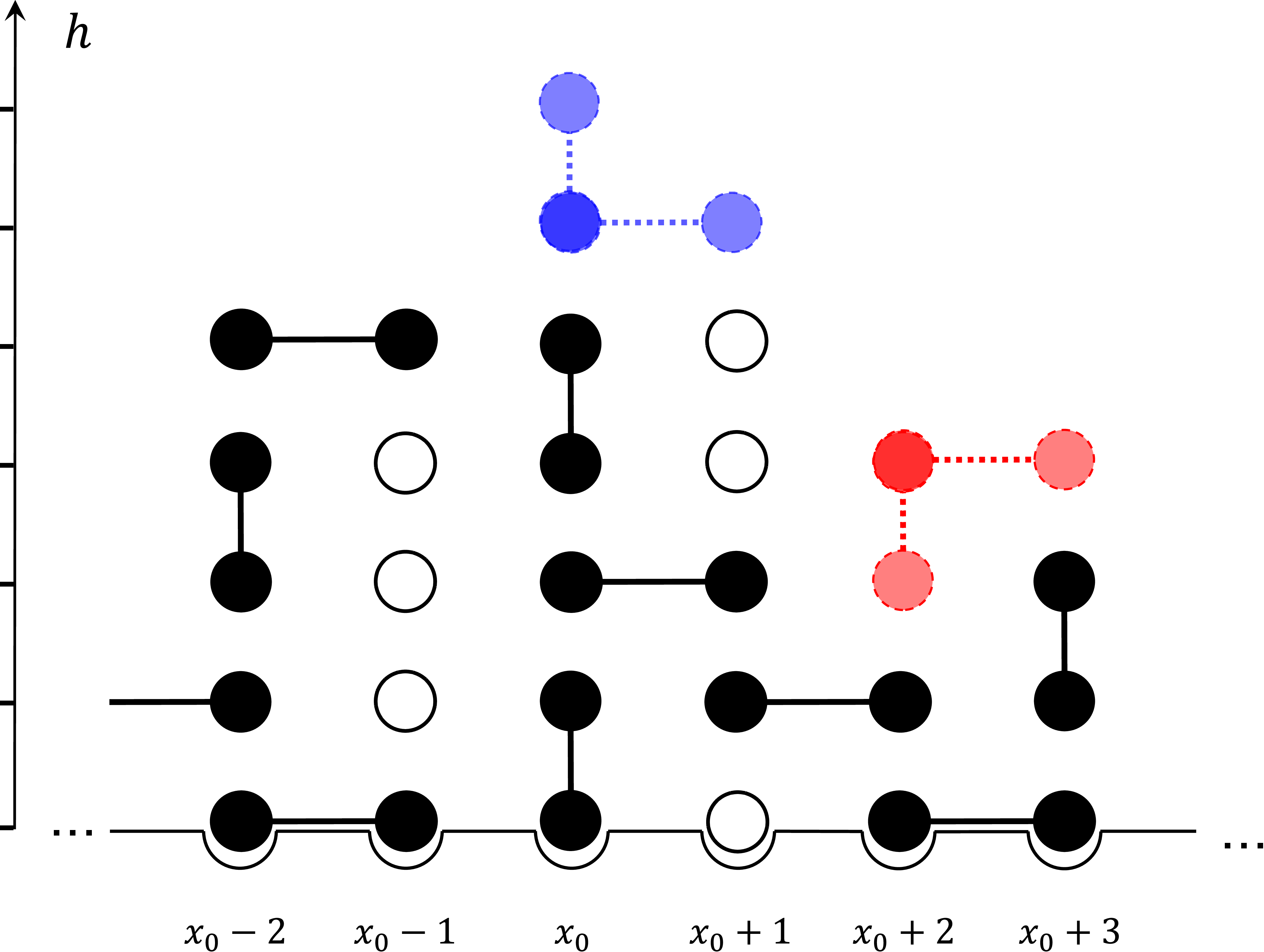} 
    
    \caption{Schematic representation of the multilayer adsorption model at an intermediate stage of the growth process using two dimers, red and blue, considering the different possibilities of adsorption (see text for more details). Blocking scenarios arise due to the presence of overhangs.}
    \label{fig:sketch}
\end{figure}

We implement the dynamics in the following way: at each instant of time $t$, by selecting the orientation of a dimer randomly with probability $p_0$ and $(1-p_0)$ for horizontal and vertical, respectively, we drop the dimer from a randomly selected position \(x_0\) onto the lattice. For a horizontal (vertical) dimer, we choose the location of one end of the dimer to be at the site:

\begin{equation}
    x_0 \in 
        \begin{cases}
           [1,L-1],& \text{if horizontal}\\
           [1,L],  & \text{if vertical}
        \end{cases}
\end{equation}
then, the dimer is placed at the sites: 
\begin{equation}
    \begin{cases}
       ([x_0, x_0 + 1],h_{max}),& \text{if horizontal}\\
       (x_0,[h_{max}, h_{max}+1]),  & \text{if vertical}
    \end{cases}
\end{equation}
where \(h_{max}\) is the maximum height of the growing structure at the corresponding columns. In Fig.\ \ref{fig:sketch}, we display a schematics of the deposition algorithm. In this sketch, we consider describing the situation of depositing two new dimers (blue and red) at two different instants of time, after depositing several dimers (painted in black). It is important to mention that first, we select the orientation (which does not change during the fall of the dimer) and then the column at position $x_0$. 

Note that, if the chosen orientation for the blue dimer is horizontal, then two voids will be created with coordinates $(x_0+1, 4)$ and $(x_0+1, 5)$, shown with an empty circle, that will never be filled due to the blockage created by the blue dimer. On the other hand, depending on the orientation drawn for the red dimer, the height of the first monomer will be $h=3$ (vertical) or $h=4$ (horizontal). In the latter case of horizontal orientation for the red dimer, the site $(x_0+2, 3)$ will remain unoccupied during the entire dynamics.

\begin{figure*}[t]
\centering
\begin{tabular}{c c c}
\includegraphics[trim={1.5cm 0.8cm 1.2cm 0.1cm},clip,width=0.66\columnwidth]{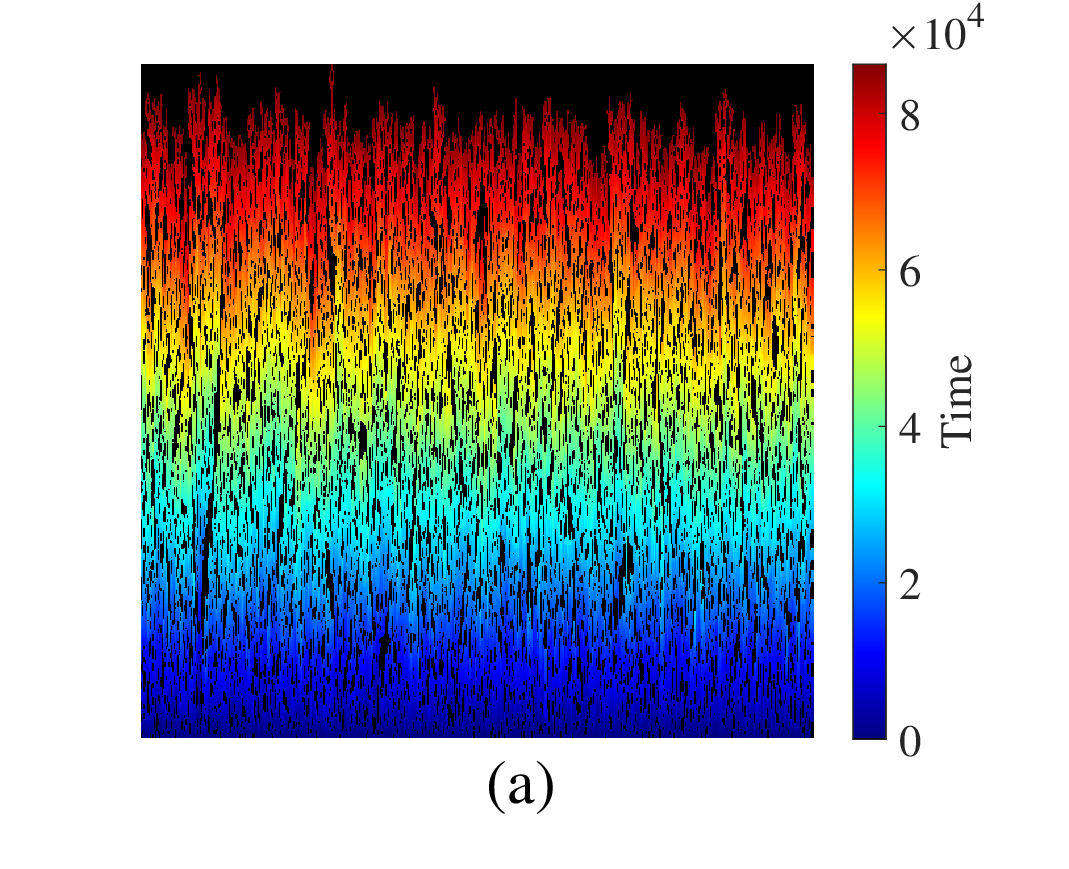} &
\includegraphics[trim={1.5cm 0.8cm 1.2cm 0.1cm},clip,width=0.66\columnwidth]{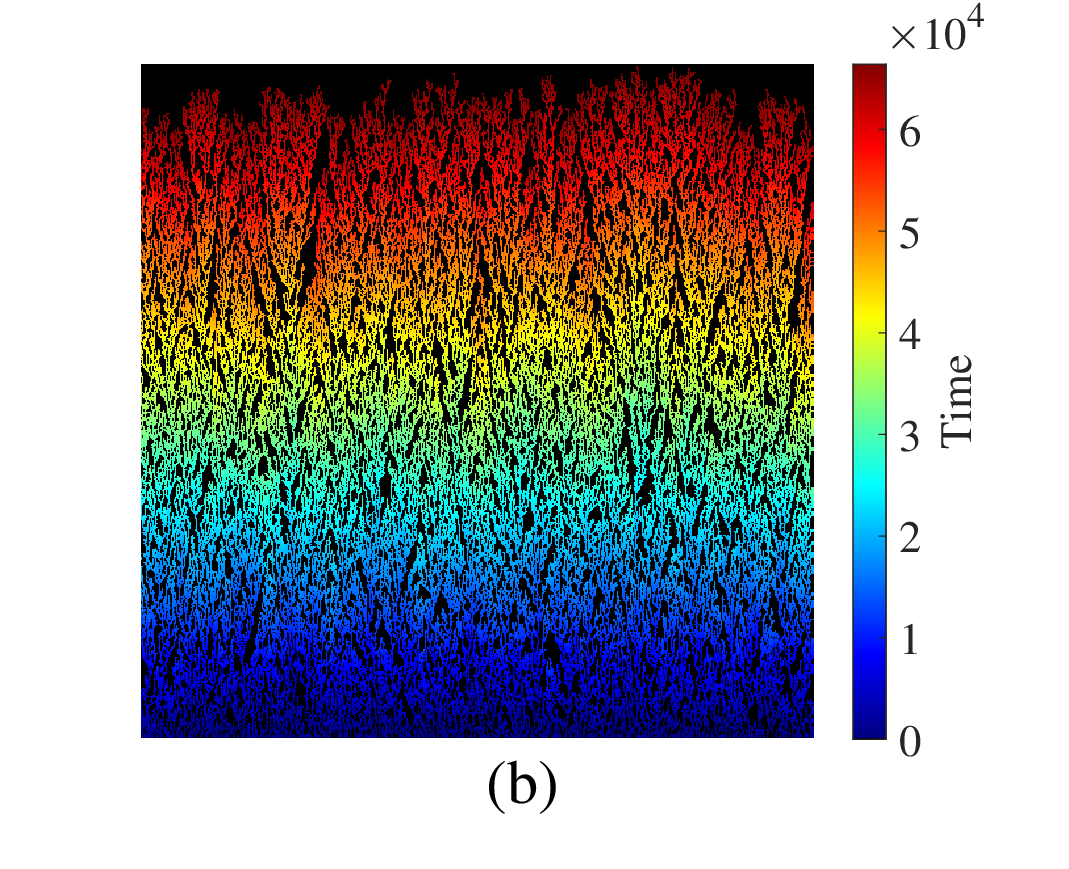} &
\includegraphics[trim={1.5cm 0.8cm 1.2cm 0.1cm},clip,width=0.66\columnwidth]{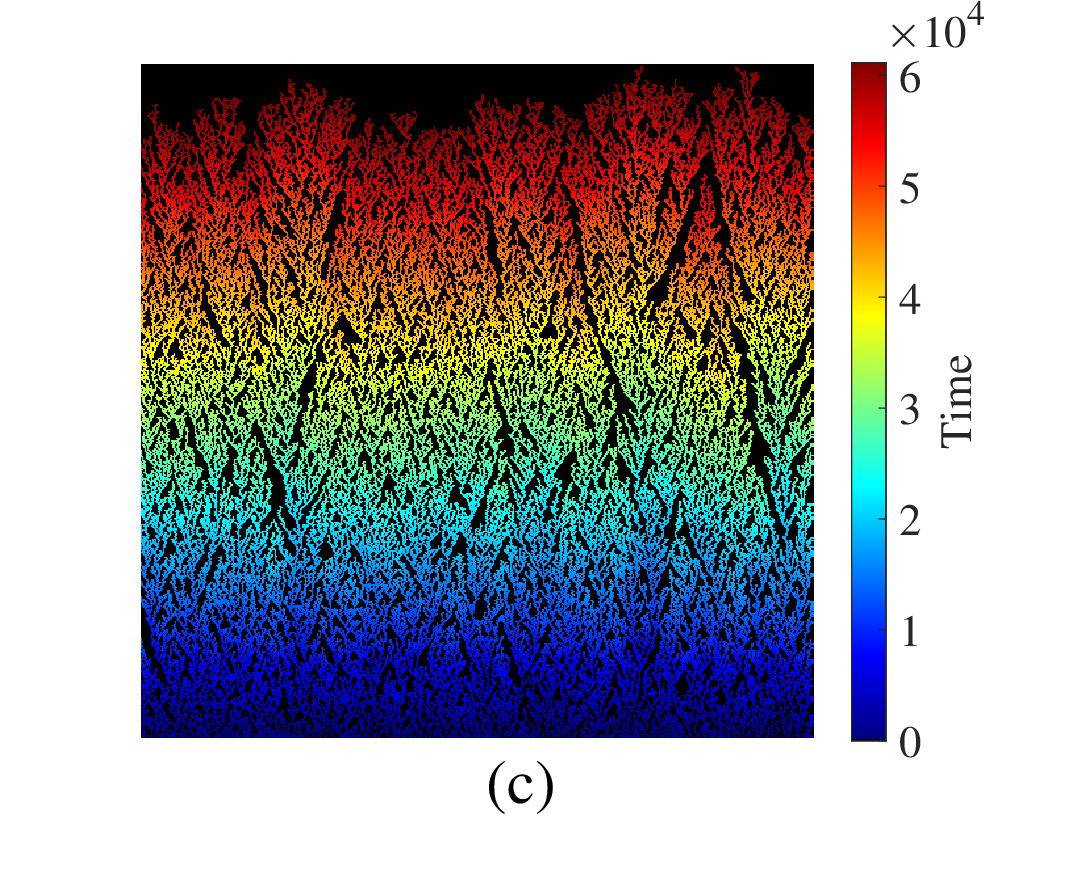}
\end{tabular}
\caption{Typical multilayer packing configurations of the model studied in the present article for $L=H_{max}=512$, and for selection probabilities \(p_0 = 0.1, 0.5, 0.9\) from (a) to (c), respectively. The dimers are colored according to their time of deposition encoded in the color bar on the right side of each figure.}
\label{fig:varp0}
\end{figure*}
\begin{figure}[b]
    \centering  
    \includegraphics[width=0.6\columnwidth]{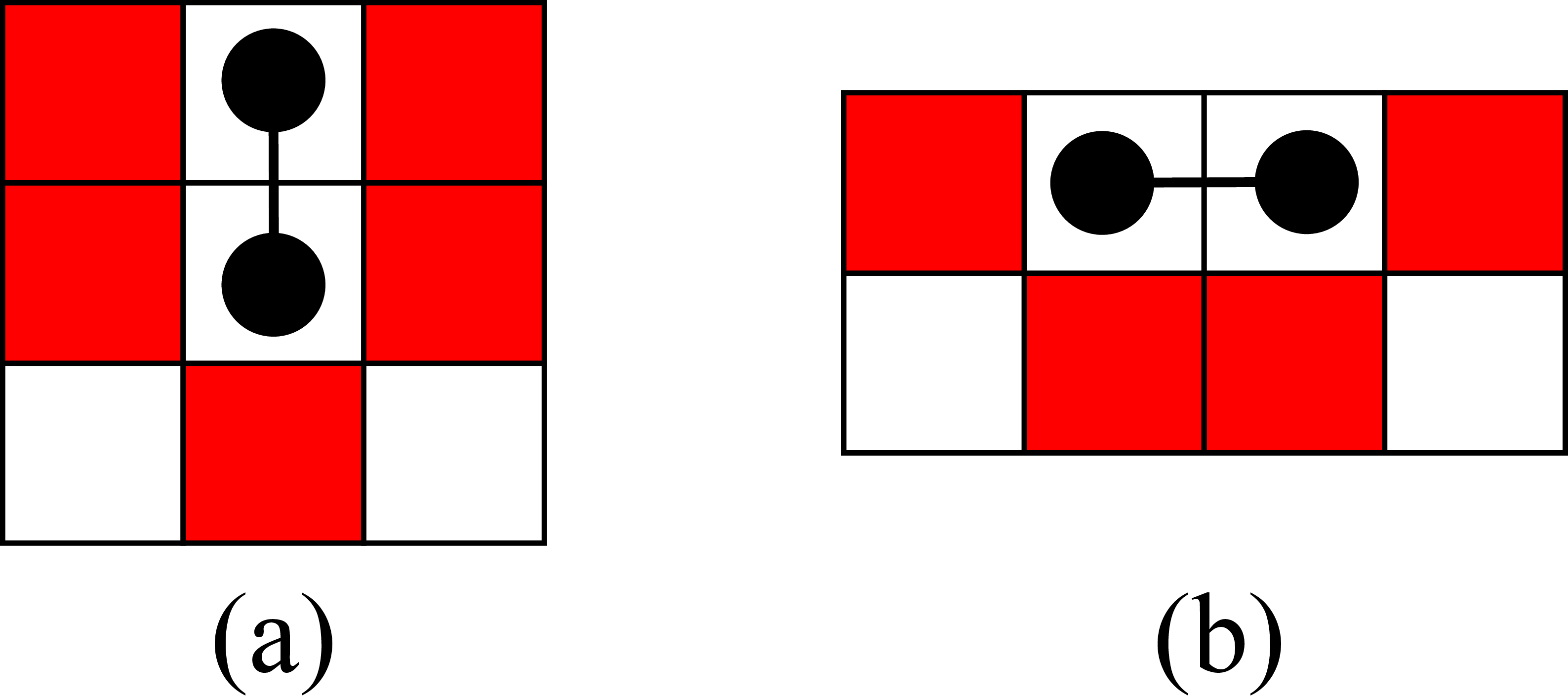}
    %\includesvg[width=0.6\columnwidth]{neigbo.svg} 
    \caption{Neighborhood connectivity criterion of the incoming dimer aligned vertically (a) or horizontally (b). The red square represents the connected neighboring sites. If there is only one label among the neighbors, then the two sites occupied by the dimer inherit that label. For different labels in the neighborhood of the dimer, a find-union procedure is applied for assigning the label to the two sites occupied by the dimer and the clusters are joined. }
    \label{fig:neig}
\end{figure}

Summarizing, the dynamics considered here always allows adhesion on top of two different dimers or on top of a single dimer, blocking the incoming dimers that arrive from the top to the bottom on the growing surface due to the presence of overhangs (screening effects~\cite{Bartelt}). At any stage, the maximum height of the entire growing structure along the vertical direction is denoted by $H_{max}$. The simulation stops when $H_{max}$ reaches a desired value. For the analysis of the critical behavior of the system, this is set at its minimum value when a percolating cluster is formed.  
Figure \ref{fig:varp0} shows typical configurations of the growing structure for different values of $p_0$.

These images suggest that the multilayer packing structure of dimers results in dendritic structures with a complex pore structure that is dependent on the parameter $p_0$. With increasing $p_0$, porous space increases and the system becomes less densely packed due to the increased screening effects from horizontal dimers. Their morphologies are basically dominated by the contribution of the complex geometry of the internal region of the structure and of the boundary, known as the active region (to be examined in subsection \ref{sect:Interface}). Both have very interesting fractal properties. It should be noted that even if the morphology of the formation might be different due to the orientational anisotropy of the dimers, the roughness exponent associated with the growing front is the same as the ordinary Ballistic deposition model.

In this work, the focus of the analysis is based on the physical properties of the percolation cluster, a cluster being a set of occupied sites connected through their nearest neighbors. To identify the percolation cluster, we combine the deposition algorithm discussed above with the Hoshen–Kopelman algorithm~\cite{Hoshen1976}. This allows the labeling of the dimers at each deposition step. To determine when the percolation cluster is created it was used the following trick: the first and last columns of the deposition grid are assigned two different labels. During the deposition process, clusters of interconnected dimers are created with different labels. The assignment of the labels is carried out following the neighborhood connectivity criterion adopted in Fig.\ \ref{fig:neig} for vertical (a) and horizontal (b) orientation of the incoming dimer. 

When the labels of the first and the last column of the deposition grid are the same, it means that the percolation cluster has emerged. We have also used the Burning algorithm~\cite{Herrmann_1984} to calculate the percolation threshold.

\section{Results and discussion} \label{sect:Results}
\subsection{Percolation} \label{sect:Percolation}

\begin{figure}[t]
\includegraphics[width=1\columnwidth]{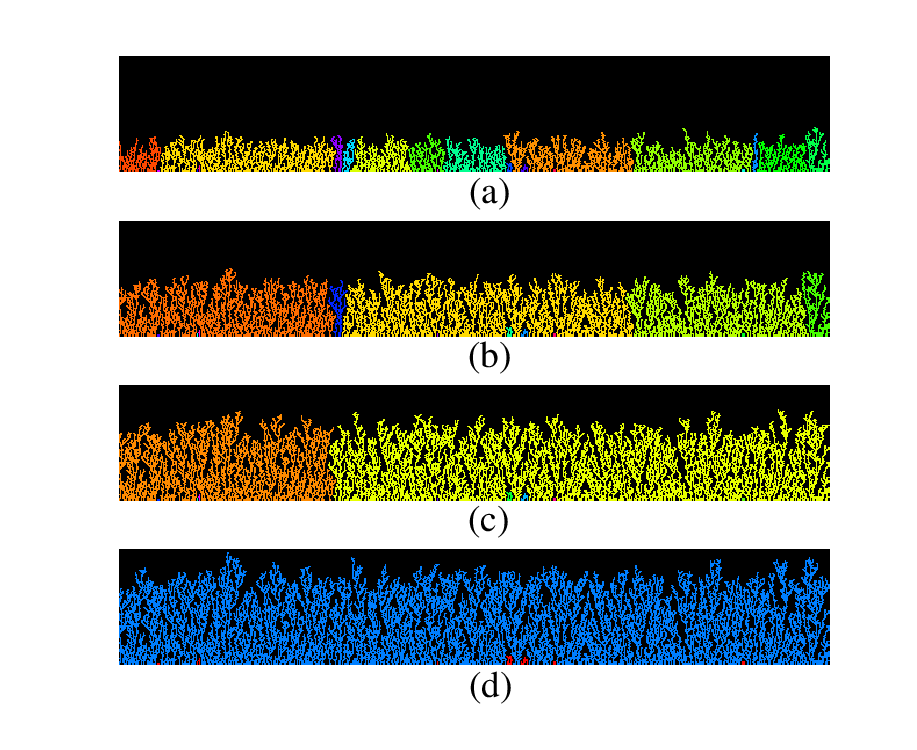}
\caption{Snapshots of the growing structure on a lattice of size $L=612$ for $p_0=0.7$ at different instants of time $t$ = 4408 (a), 7466 (b), 10330 (c), and 12343 (d). The corresponding values of $H_{max}$=40, 60, 80, and 98, respectively. Different colors represent different clusters.}
\label{fig:culsters}
\end{figure}

The focus of this subsection is to study the percolation properties of the growing structure formed due to particle deposition. Note that the entire growth process can be classified into three stages in time (see Fig.\ \ref{fig:culsters}): (i) Initial stage: increase in the number of isolated clusters of dimers connected through their nearest neighbors and their size growth with time, (ii) Intermediate stage: merging of growing clusters of different sizes, and (iii) Final stage: size growth of only one single cluster. At this stage, there may also be other clusters, but their growth is stopped, as a new incoming dimer cannot penetrate the deep interior of the formed structure due to the screening effect.

\begin{figure} [t]
    \centering    
    \subfigure{\includegraphics[width=0.480\columnwidth]{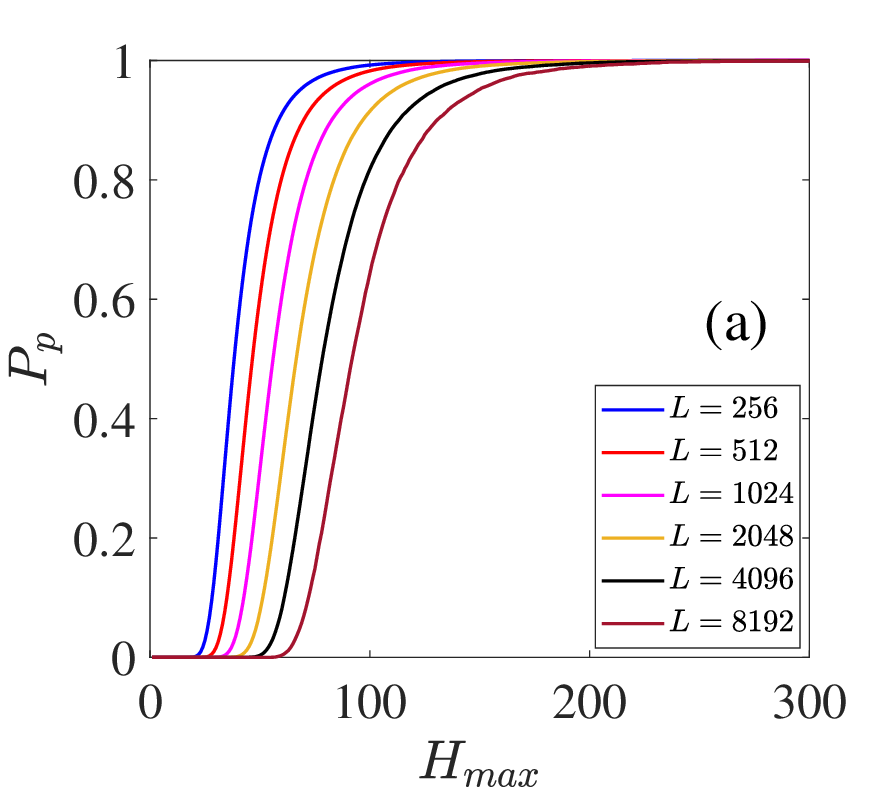}}    
    \subfigure{\includegraphics[width=0.480\columnwidth]{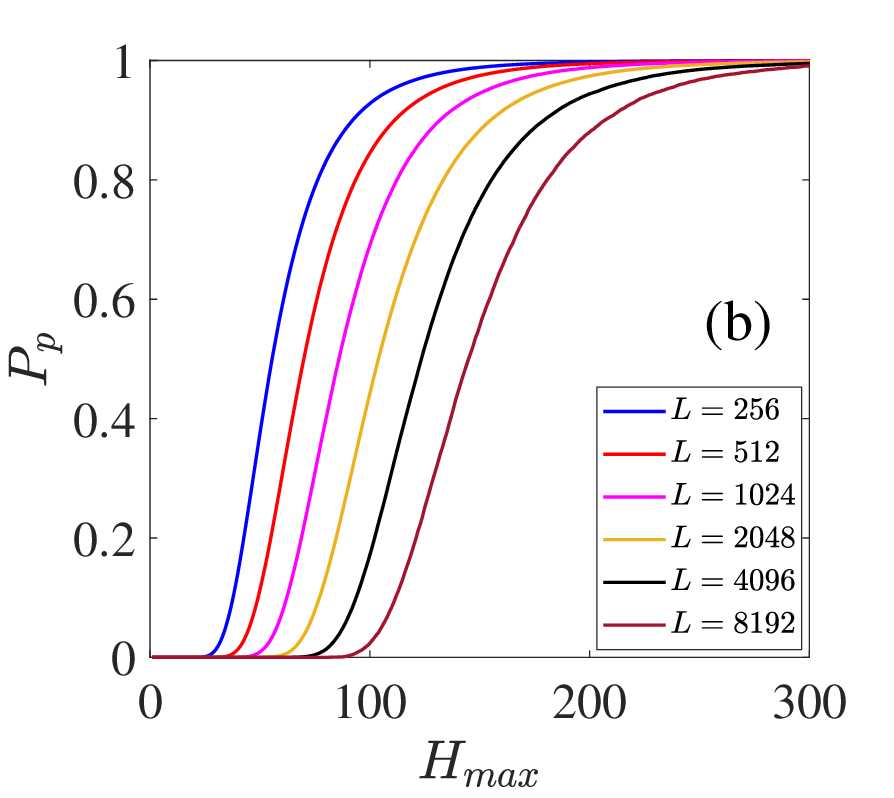}}\\    
    \subfigure{\includegraphics[width=0.480\columnwidth]{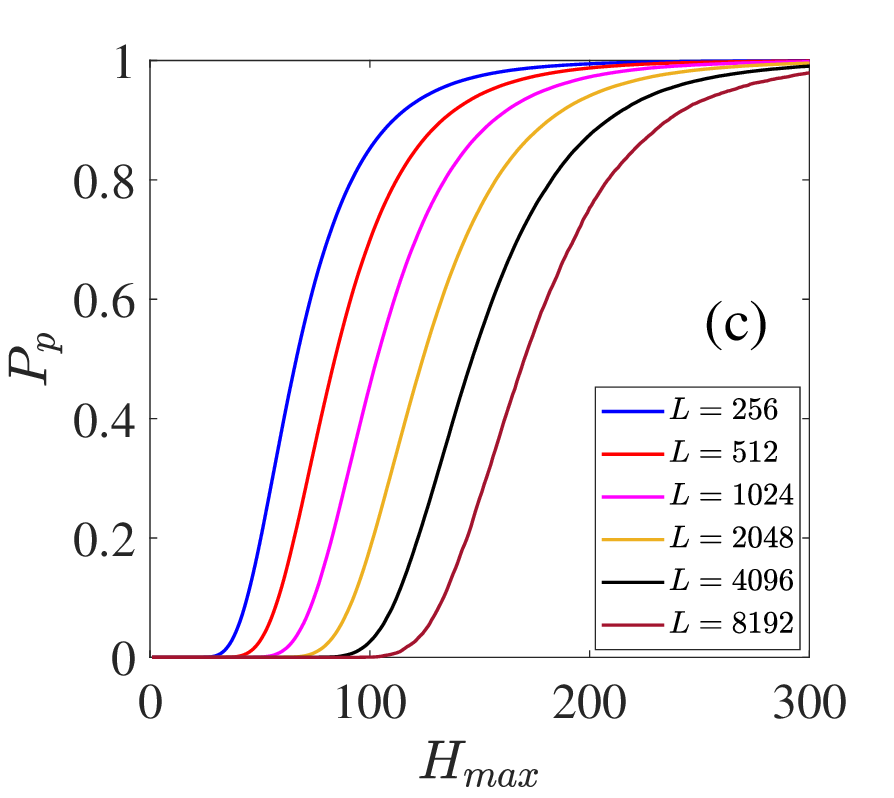}}    
    \subfigure{\includegraphics[width=0.480\columnwidth]{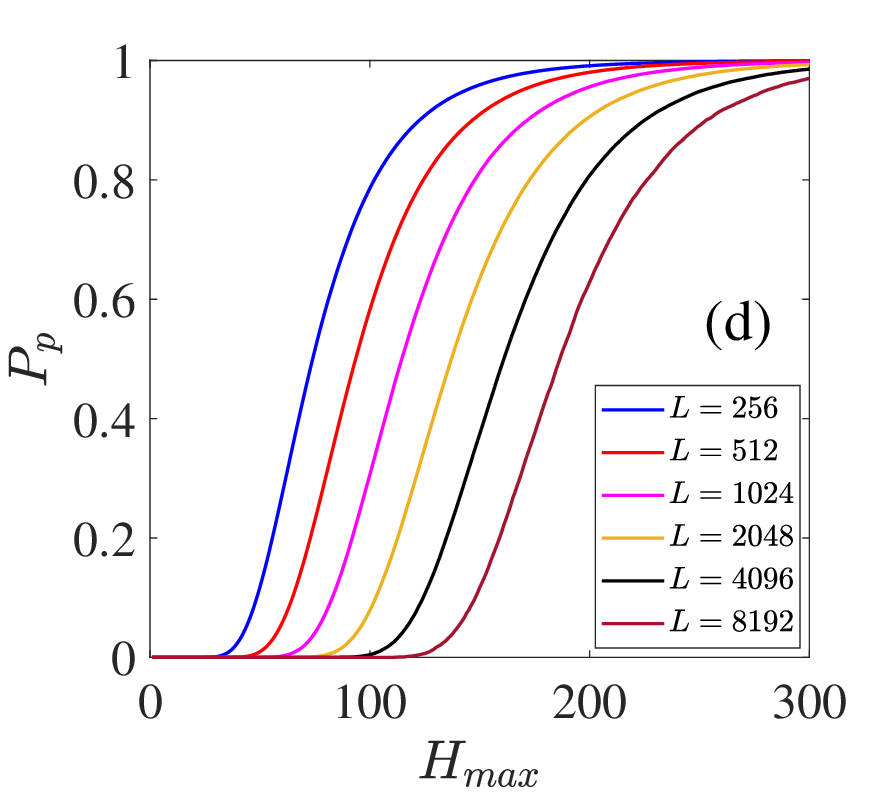}}
    \subfigure{\includegraphics[width=0.480\columnwidth]{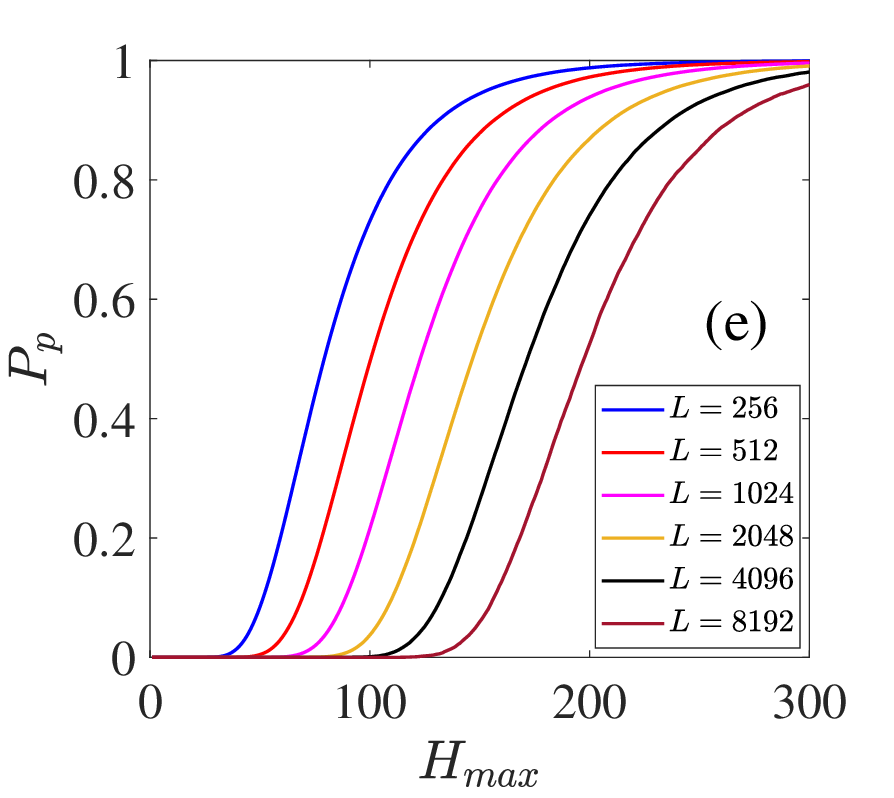}}
    \subfigure{\includegraphics[width=0.480\columnwidth]{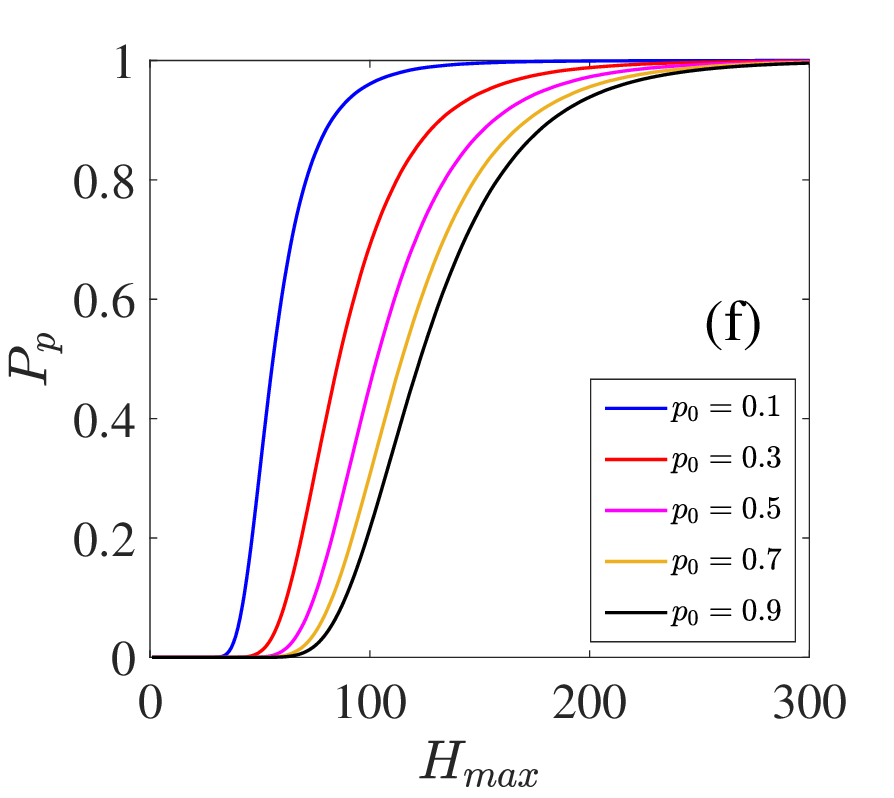}}
    \caption{Percolation probability \(P_p\) as a function of the maximum height $H_{max}$ of the multilayer growing structure for $p_0$ = 0.1 (a), 0.3 (b), 0.5 (c), 0.7 (d), 0.9 (e). (f) The plot of \(P_p\) vs. $H_{max}$ for different values of \(p_0\) using \(L=1024\), showing clearly its dependence on $p_0$. The results are based on averages over $10^6$ (for the first two smaller systems) to 13750 (for the largest system) independent samples. For $L=1024$, we consider (at least) $2.2 \times 10^5$ samples.}
    \label{fig:Pp}
\end{figure}

When the $H_{max}$ of the entire structure is sufficiently high, there exists a spanning path from the left to the right of the system through the largest cluster at the final stage, whereas such a spanning cluster is absent at the initial stage. At the intermediate stage of the cluster merging process, as $H_{max}$ gradually increases, a percolation transition occurs when such a spanning cluster first appears between the left and right boundaries of the system at a critical value of $H_{max}=H_c(L)$. Figs.\ \ref{fig:Pp}(a-e) display the left to right spanning probability \(P_p\) of the system as a function of \(H_{max}\) on a lattice of linear size $L$ ranging from 256 to 8192 for five different values of $p_0$. The curves clearly depend both on $L$ and $p_0$.

\begin{figure}[t]
\centering
\includegraphics [width=1\columnwidth]{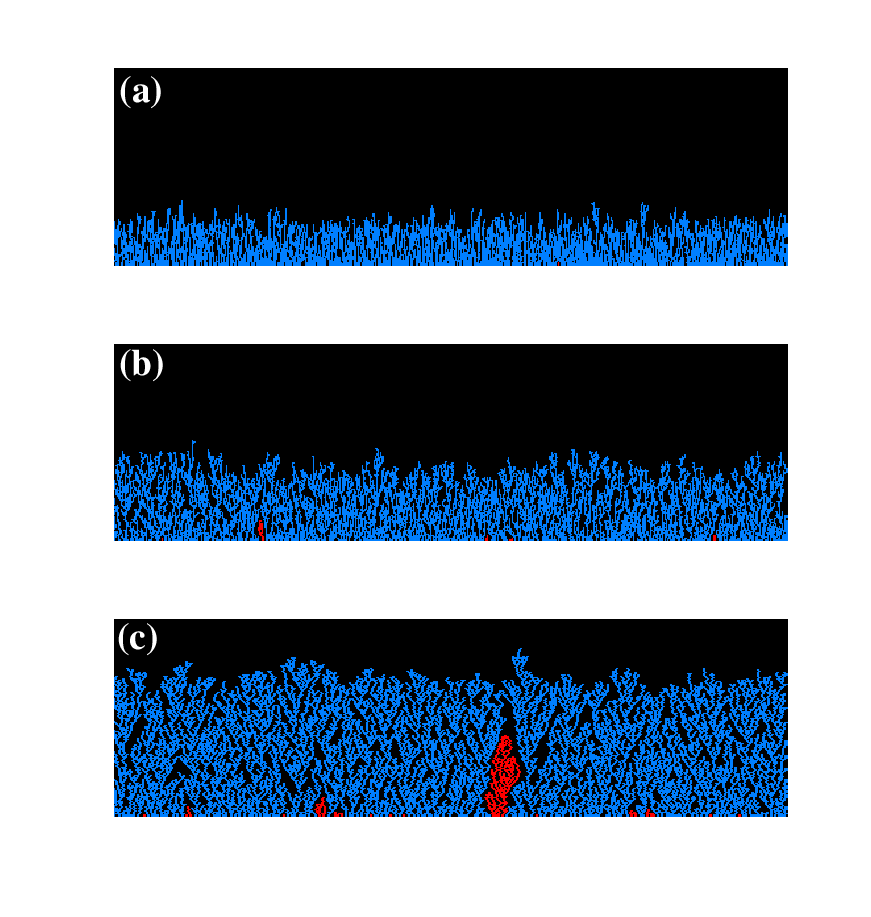}
\caption{\label{fig:wide}The growing structure on a lattice of size $L=512$ at the percolation threshold for $p_0$=0.2 (a), 0.5 (b), and 1.0 (c). The critical height at which percolation transition occurs grows with increasing the value of $p_0$. The blue and red colors represent the percolating cluster and all other isolated clusters, respectively.}
\label{fig:Hcincrease}
\end{figure}

Notice that for a specific value of $L$, the sharp rise of the curve for \(P_p\) shifts to the higher value of $H_{max}$ with increasing the value of $p_0$ (see Fig.\ \ref{fig:Pp}(f)). It indicates that the presence of vertical dimers promotes percolation. 
Qualitatively, such a behavior can be understood in the following way: for small $p_0$ the interior of the percolation cluster is more homogeneous and all the deposited dimers belong to the cluster. When $p_0$ is increased, due to the increased screening effect from horizontal dimers, at the base of the packing structure isolated clusters are formed that will never help the growing structure to establish a spanning path (see Fig.\ \ref{fig:Hcincrease}). This implies that more time (or more incoming dimers) is required for the structure to grow to obtain a spanning path. Correspondingly, the height of the percolating cluster at the percolation transition increases. Finally, it is important to mention that a finite size scaling was done on the $P_p(L)$ curves using the curve collapse method described in~\cite{Palacios2022}, but no universal exponents were found when $p_0$ is varied.

Numerically, the precise value of the percolation threshold $H_c(L)$ for a given value of $L$ and $p_0$ is determined using the bisection method~\cite{Kundu2017,Kundu2018}. We start with a pair of values of deposition time $t^{\text{hi}}$ and $t^{\text{low}}$ such that a spanning cluster exists at $t=t^{\text{hi}}$, but not at $t=t^{\text{low}}$. This interval is then successively bisected and checked if there exists a spanning cluster using the Burning algorithm~\cite{Herrmann_1984} until $t^{\text{hi}}-t^{\text{low}}=1$. At this stage, the corresponding height of the entire structure at $t=t^{\text{hi}}$ defines the critical height for a given run. Note that, we initially stored the sequence of dimer deposition on the lattice sites up to $t=t^{\text{hi}}$ and used this sequence during the iterative process. By repeating the entire procedure for a large number of independent runs and averaging the corresponding critical height values the percolation threshold $H_c(L)$ is obtained.

\begin{figure}[t]
    \centering    
    \subfigure{\includegraphics[width=0.494\columnwidth]{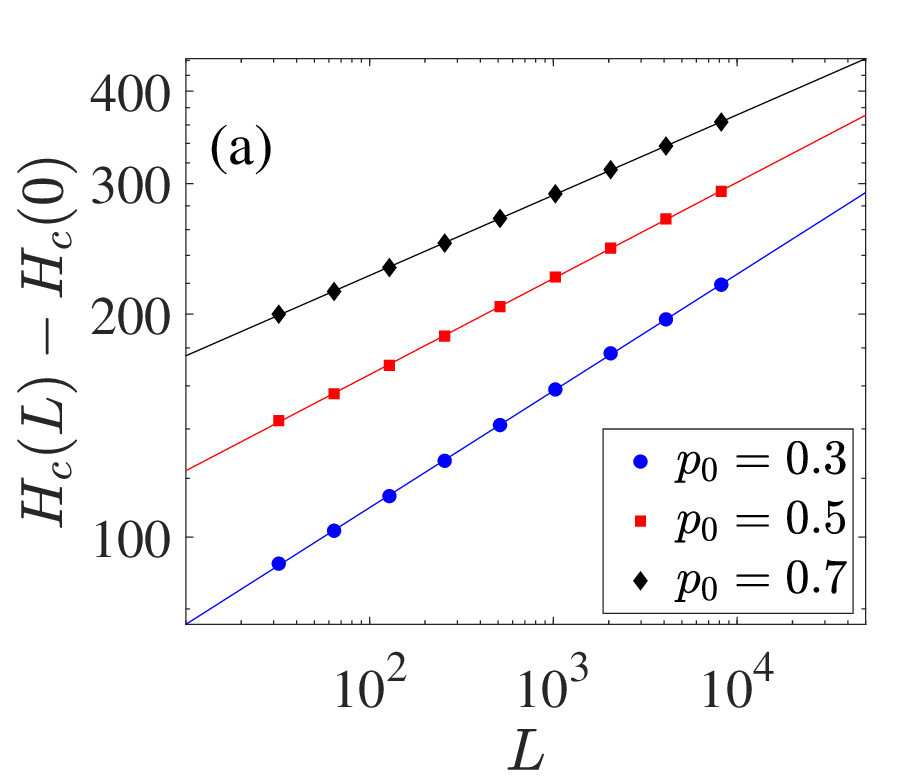}}    
    \subfigure{\includegraphics[width=0.494\columnwidth]{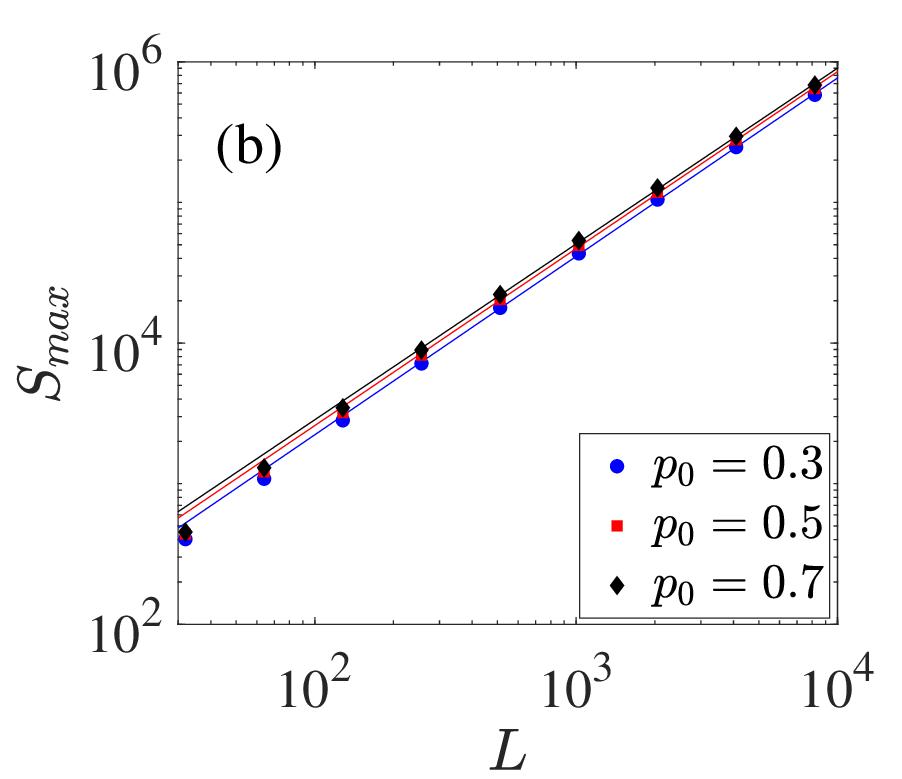}}    
    \caption{Size-scaling of (a) \(H_c(L)\), and (b) \(S_{max}(L)\) for \(p_0 = 0.3, 0.5,\) and \(0.7\). The solid line represents the fit using Eq.\ \ref{eq:powsat}. The number of samples used to find out these results are indicated in Fig.\ \ref{fig:Pp}.}
    \label{fig:scalingHcandSmax}
\end{figure}

The dependence of $H_c(L)$ on $L$ is exhibited in Fig.\ \ref{fig:scalingHcandSmax}. It appears that the data points are consistent with the following functional form:
\begin{equation}
    H_c(L) = A L^{\nu} + c,
    \label{eq:powsat}
\end{equation}
where $c$ and $A$ are the fitting constants and $\nu>0$. We find that the exponent value $\nu$ has a slight dependence on \(p_0\) values: \(\nu(p_0=0.3) = 0.158\pm0.004\), \(\nu(p_0=0.5) = 0.130\pm0.004\) and \(\nu(p_0=0.7)=0.108\pm0.003\).

In Fig.\ \ref{fig:scalingHcandSmax}(b), the finite-size scaling of the size of the percolation (spanning) cluster is presented. In this case, we fit the data using a power-law similar to Eq.\ \ref{eq:powsat} but with \(c=0\). It was verified that for sufficiently large \(L\), \(S_{max}(L)\) scales as: $S_{max}(L) \sim L^\alpha$ with \(\alpha(p_0=0.3) = 1.268 \pm 0.006\), \(\alpha(p_0=0.5) = 1.258 \pm 0.007\) and \(\alpha(p_0=0.7) = 1.250 \pm 0.007\). Note that irrespective of the value of $p_0$ we obtain nearly the same exponent value $\alpha$ associated with the variation of \(S_{max}(L)\).

\subsection{Fractal and diffusion dimension of the percolation cluster} \label{sect:Fractal}

This subsection will be dedicated to the study of the fractal and diffusion properties associated with the percolation clusters generated with our model. It should be noted that in this case, we cannot use the usual way of calculating the fractal dimension in which the scaling exponent of the mass (or area) of the percolation cluster with \(L\) is exactly the fractal dimension. 

This is an anisotropic system where the height \(H_c\) scales non-trivially with \(L\). That is why one should use some alternative method. To obtain their corresponding fractal dimensions a method based on the density autocorrelation function was chosen; the averaged density of occupied sites within neighborhoods of radius \(r\) centered in all points belonging to the cluster is first calculated. Afterward, a log-log plot (not shown) of density against distance \(r\) then yields a straight line with slope \(D_f-D\), where \(D\) is the dimensionality of the space (2 in this case). For each value of $p_0$, the fractal dimension for each $L = 128, 256, 512,$ and $1024$ is obtained by averaging the data over 16000, 2800, 400, and 100 replicas, respectively. The diffusion exponent \(D_w\) \cite{Havlin1987} belonging to the percolation cluster was also calculated. For this, random walks are generated starting at a random occupied site of the percolation cluster, allowing to walk upto time steps \(t=100, 200, 500, 1000, 5000, 10000, 100000\), and then, determining in each case the average of the square of the distance between the starting and ending point of the walk \(\langle r^2 \rangle \). The average \(\langle r^2 \rangle \) for each value of \(t\) is taken with different number of replicas, i.e., \(N_{rep} = 73000, 44000, 20000, 10000, 2000, 1000, 100\), respectively. Finally, the fractal dimension \(D_w\) of the walk is defined by the scaling relation \(\langle r^2 \rangle \sim t^{2/D_w}\), which
reduces to the known result \(\langle r^2 \rangle \sim t\) for Brownian walkers (\(D_w=2\)).

\begin{figure}[t] 
\centering  
 \includegraphics[width=1\columnwidth]{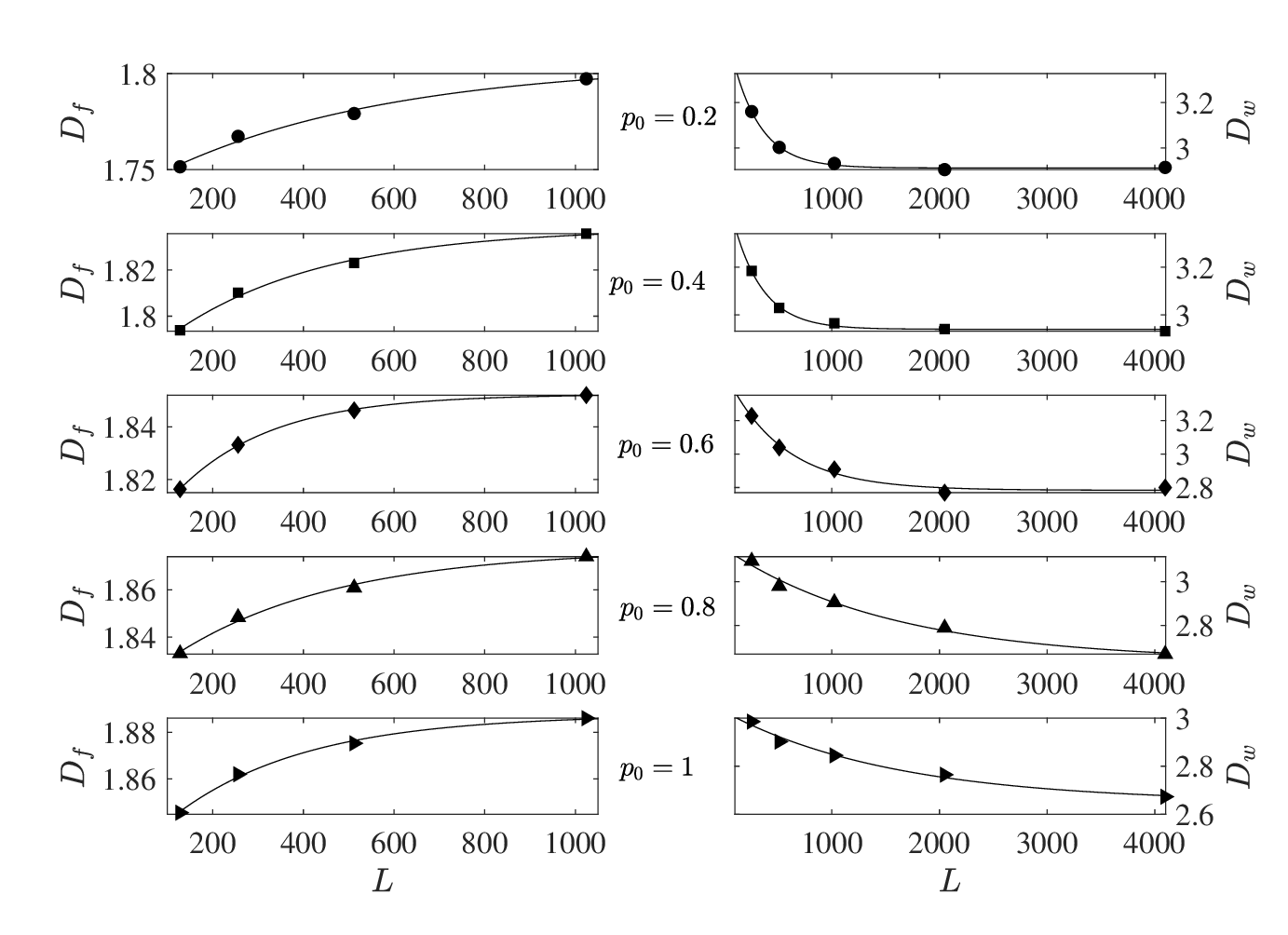}
    \caption{In the left column we show the fractal dimension \(D_f\) of the percolation cluster for different values of the probability \(p_0\) and for different sizes \(L\) of the substrate. In the right column, we show the diffusion exponent. The solid line in all the plots represents the fit of the data using Eq.\ \ref{eq:expmodel}.
    In all cases, the error bar (variance) is much smaller than the size of the symbols.}
    \label{fig:fractal_and_diffusin}
\end{figure}

The behavior of the fractal dimension \(D_f\) as a function of \(L\), for different values of the probability \(p_0\) and the corresponding diffusion exponent \(D_w\) are shown, respectively, in the left and right columns of Fig. \ref{fig:fractal_and_diffusin}.

Asymptotic analyses of \(D_f(L \to \infty)\) and \(D_w(L \to \infty)\) were done for all the curves shown in Fig.\ \ref{fig:fractal_and_diffusin}. The data of \(D_x(L)\) (for $x = f, w$) for the different values of \(p_0\) was fitted using the following exponencial form: 
\begin{equation}
D_x(L) = a e^{-bL}+c.
 \label{eq:expmodel}
\end{equation}
The result of the fit is shown in Fig.\ \ref{fig:fractal_and_diffusin} using a solid line in each case. Table \ref{table:1} shows the values of the fitted asymptotic coefficients \( c=D_x(L \to \infty)\) valid in the thermodynamic limit, for each value of \(p_0\).
 
One may note that the asymptotic value of \(D_w(L\to\infty)\) closely matches its classical percolation value of \(D_w = 2.8784\) \cite{Grassberger1999} when \(p_0\) is close to 0, however, in this regime, \(D_f\) is more distant from its classical percolation value of 91/48 that is approached here in the limiting case of $p_0$ close to unity. Notably, Ben-Avraham and Havlin \cite{Ben-Avraham_1982,ben-avraham_2000} reported \(D_w = 2.7 \pm 0.1\) for classical 2D percolation, which is close to our obtained values for \(p_0>0.6\) (see Table \ref{table:1}).

\begin{table}[t]
\centering
\begin{tabular}{|p{1.0cm}|p{3.0cm}|p{3.0cm}|} 
 \hline
 \(p_0\) & \(D_f\) & \(D_w\) \\ [0.5ex] 
 \hline
 0.2   &  \(1.806 \pm 0.037\)  & \(2.912 \pm 0.099\)\\
0.4   &  \(1.838 \pm 0.027\)  & \(2.939 \pm 0.124\)\\
0.6   &  \(1.852 \pm 0.010\)  & \(2.783 \pm 0.137\)\\
0.8   &  \(1.877 \pm 0.024\)  & \(2.638 \pm 0.065\)\\
1.0   &  \(1.888 \pm 0.024\)  & \(2.649 \pm 0.050\)\\ [1ex] 
 \hline
\end{tabular}
\caption{Values of fractal dimension (\(D_f\)) and diffusion exponent (\(D_w\)), obtained from the asymptotic analysis using Eq.\ (\ref{eq:expmodel}) and representing the approximate value of such quantities in the thermodynamic limit.}
\label{table:1}
\end{table}

\subsection{Electrical conductivity} \label{sect:cunct}
The Frank and Lobb algorithm~\cite{Frank1988} was used for finding the conductivity between the left and right boundaries of the two-dimensional structure that results from the dimer deposition process. The same approach described in~\cite{Palacios2022} was followed. The calculations of the conductivity \(\sigma\) were performed each time until the multilayer reaches a given height. For each given value of \(H_{max}\), the computer experiments were repeated 1200 times.

\begin{figure}[t] 
\centering  
\subfigure{\includegraphics[width=0.494\columnwidth]{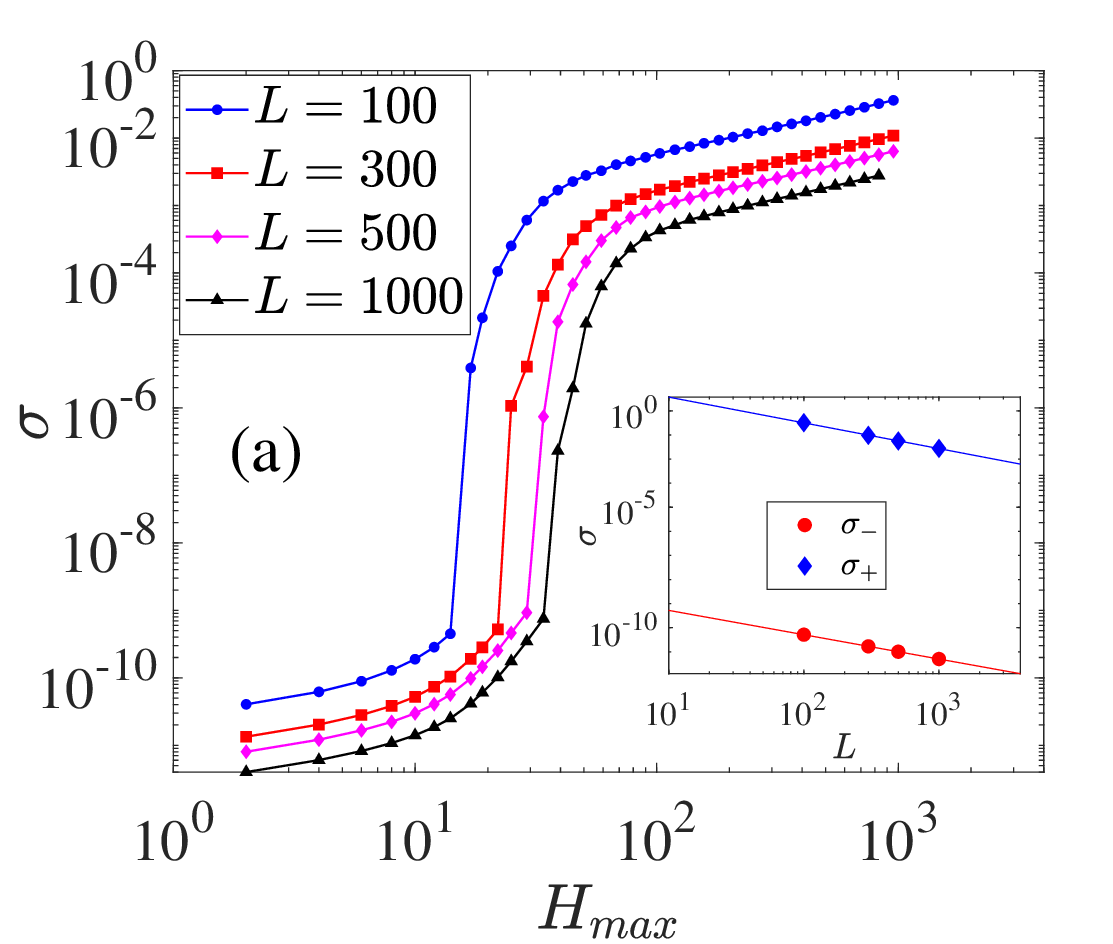}}
\subfigure{\includegraphics[width=0.494\columnwidth]{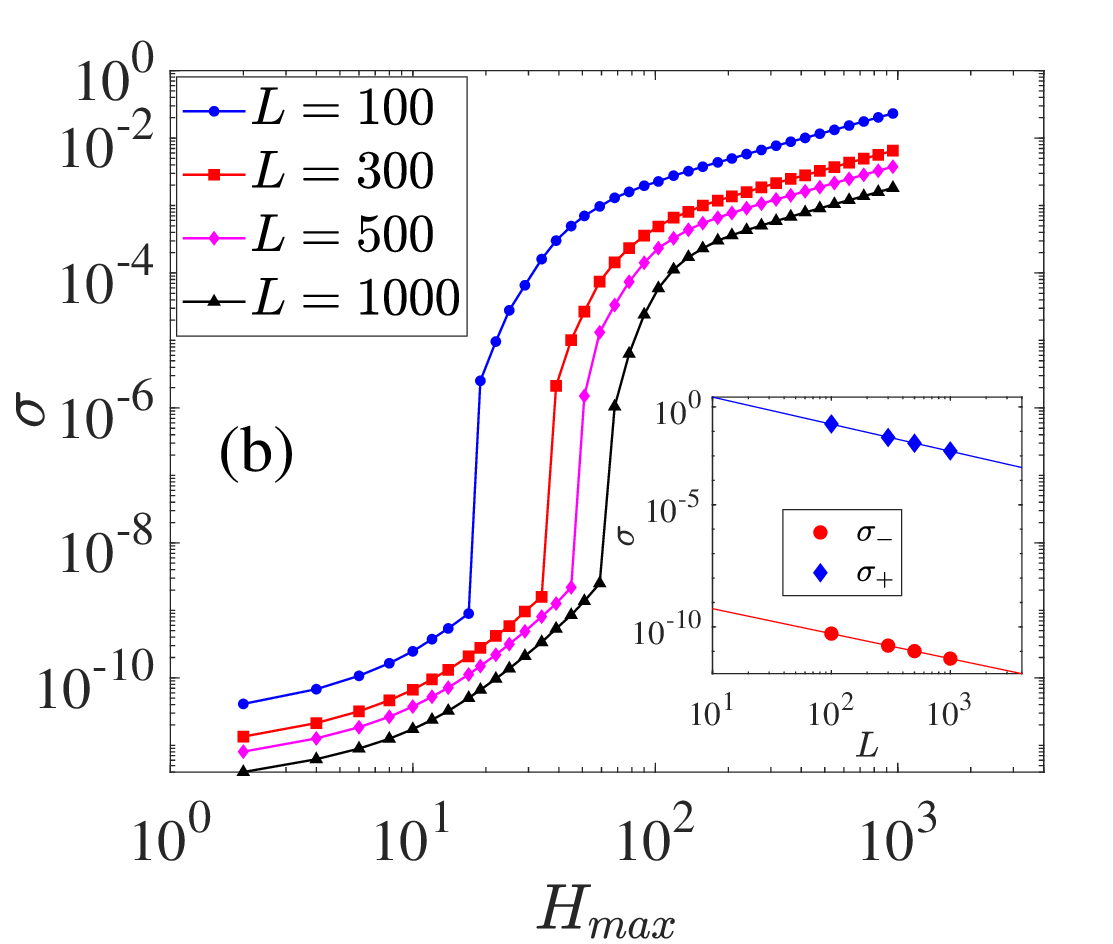}} 
    \caption{Mean conductivity as a function of \(H_{max}\) for \(L=100, 300, 500, 1000\) and for \(p_0=0.2\) (a) and \(p_0=0.8\) (b). Inset shows the scaling analysis of conductivity \(\sigma_{+(-)}\) after (before) the percolation transition. In the main plot and in the inset, the size of the symbols is larger than the corresponding error bars.}
    \label{fig:scalingsigma}
\end{figure}

In Fig.\ \ref{fig:scalingsigma}, we show the behavior of the mean conductivity as a function of \(H_{max}\) for different values of \(L\) and for two values of \(p_0\).

As is known, the percolation phase transition occurs when a system undergoes a transition from an insulating state to a conducting state. In general, there exists a power-law scaling relation between the electrical conductivity \(\sigma\) and the linear size \(L\) of the structure near the percolation threshold through two different scaling relations for \(\sigma_-\) (conductivity before the percolation transition) and \(\sigma_+\) (conductivity after the percolation transition)~\cite{Cherkasova2010, Tarasevich2017},i.e.,
\begin{equation}
\sigma_{\pm} \sim L^{-\gamma_{\pm}}.
 \label{eq:expmodel1}
\end{equation}
In Table \ref{table:2}, we list the values of the scaling exponents \(\gamma_-\) and \(\gamma_+\) for the same values of \(p_0\) used in Table \ref{table:1}.

\begin{table}[b]
\centering
\begin{tabular}{|p{1.0cm}|p{3.0cm}|p{3.0cm}|} 
 \hline
 \(p_0\) & \(\gamma_-\) & \(\gamma_+\) \\ [0ex] 
 \hline
0.2   &  \(1.010 \pm 0.003\)   &  \(1.069 \pm 0.019\)\\
0.4   &  \(1.012 \pm 0.003\)   &  \(1.073 \pm 0.018\)\\
0.6   &  \(1.016 \pm 0.004\)   &  \(1.091 \pm 0.026\)\\
0.8   &  \(1.022 \pm 0.005\)   &  \(1.107 \pm 0.031\)\\
1.0   &  \(1.017 \pm 0.013\)   &  \(1.129 \pm 0.116\)\\ 
 \hline
\end{tabular}
\caption{Scaling exponents $\gamma_{\pm}$ of the conductivity before ($-$)  and after ($+$) the percolation transition. The data is fitted using the power-law scaling relation in Eq.\ (\ref{eq:expmodel1}).}
\label{table:2}
\end{table}

The first interesting result is that for $H_{max}<H_c$, i.e., below the percolation transition, the system presents a universal behavior, which is reflected through a constant value (equal to 1) of the scaling exponent of \(\sigma_-\), regardless of the value of \(p_0\). This value for the scaling exponent implies that the system belongs to the universality class of percolation transition in two dimensions~\cite{Jang2019,Bunde1996}. The scaling exponent for \(\sigma_+\) is more interesting. The results shown in the third column of Table \ref{table:2} suggest that just above the percolation point, the critical behavior of the conductivity depends on \(p_0\). More specifically, in the post-critical regime, the decay of conductivity \(\sigma_+\) with $L$ becomes faster with increasing $p_0$, i.e., horizontal alignment induces the decrease of the conductivity for a given set of parameter values.

The electrical conductivity in amorphous disordered systems crucially depends on its own structural details. The notion of percolation theory and fractals are useful to understand this~\cite{Ben-Avraham_1982, ben-avraham_2000}. A scaling relation, as described below, can be obtained that uncovers this non-trivial dependency.

In a uniform Euclidean system, the Einstein ~\cite{Havlin1987,Iguain2022} relation says that the mean square displacement \(\langle r(t)^2\rangle\) of a random walker is proportional to time \(t\), in any spatial dimension. However, for disordered systems, this linear relationship is not valid in general, and is described by:
\begin{equation}
\langle r^2(t)\rangle \sim t^{2/D_w}.
 \label{eq:rsm}
\end{equation}

As mentioned in \ref{sect:Fractal}, the diffusion exponent assumes a value $D_w>2$ different from the classical Brownian exponent \(D_w=2\). This delay in transport is caused by the peculiarities in the spatial distribution of scattering centers in the disordered structure~\cite{Gefen1983}. Regarding geometric and structural aspects, the electrical resistance \(R\) of a material depends both on the size and the space dimensionality \(D\) and on the topology of the system as: \(R = L/(\sigma A\), where \(L\) is the distance between the electrodes and \(A\) is the cross-section area. This implies that for a \(D\)-dimensional Euclidean system (whose area \(A\) is proportional to \(L^{D-1}\)) one gets:
\begin{equation}
R \sim L^{2-D}.
\label{eq:rscal}
\end{equation}
Therefore, for homogeneous square samples the electrical resistance does not depend on its size.

From the microscopic point of view, the electrical resistance of a system
can be written in general terms as:
\begin{equation}
R \sim N/M,
\label{eq:rnm}
\end{equation}
where \(N\) is the average number of scattering centers in the system of length \(L\), and \(M\) is the number of possible scattering centers \cite{Gomes_2011}. The denominator \(M\) is proportional to the number of atoms in the system, \(L^D\), whereas the numerator \(N\) is proportional to the average time required for the electron to go through the distance between the ends of the system. From Eq.\ (\ref{eq:rsm}), it is expected that \(N\) is proportional to \(t \sim \langle r^2(t)\rangle ^{D_w/2} = (L^2)^{D_w/2}\). For a non-uniform fractal system, one can further express Eq.\ (\ref{eq:rnm}) in terms of \(L\) by assuming that the resistor has a fractal dimension \(D_f\) as: \(R \sim \frac{N}{M} \sim \frac{L^{D_w}}{L^{D_f}} \sim L^{D_w-D_f}\)  \cite{ben-avraham_2000}, or equivalently, for the electrical
conductance:
\begin{equation}
C \sim L^{D_f-D_w}. 
\label{eq:Rscaling}
\end{equation}
Then, in a two-dimensional medium and with diffusion exponent \(D_w=2\) the electrical conductance scales as \(C \sim L^{\alpha}\), where \(\alpha=D_f-D_w=0\). For a media with fractal dimension different from the diffusion exponent, it is expected that \(\alpha \neq 0\). For instance, for a percolation cluster in two dimensions, \(D_f = 91/48 = 1.8958\) and \(D_w=2.8784\) \cite{Grassberger1999}, and therefore, \(C \sim L^{D_f-D_w} = L^{1.8958 - 2.8784} = L^{-0.9826}\). 

\begin{figure} [t]
\centering 
    \subfigure{\includegraphics[width=0.80\columnwidth]{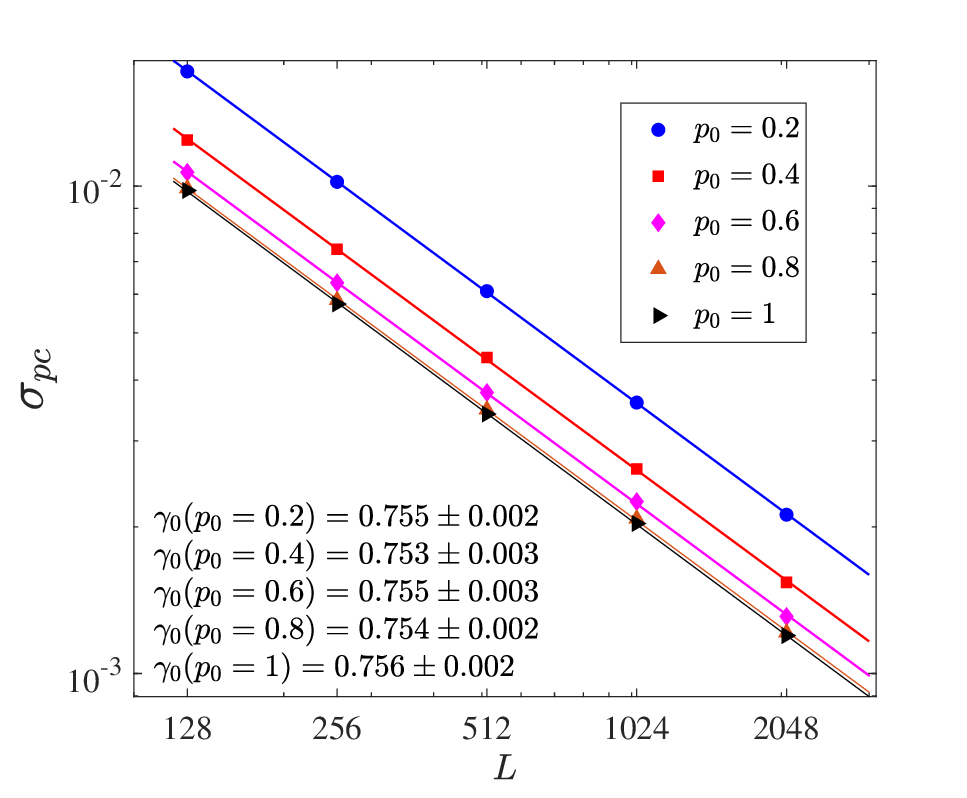}}         
    \caption{Scaling of the conductivity of percolation cluster for different values of the probability \(p_0 = 0.2; 0.4; 0.6; 0.8; 1\). The solid lines represent the fit to power law model \(\sigma \sim L^{-\gamma_0(p_0)}\) whose exponents \(\gamma_0\) are inserted in the figure.}
    \label{fig:areaesigmapc}
\end{figure}

In order to check the validity of this last result the scaling behavior of the conductivity of the percolation cluster was calculated. The results are shown in Fig.\ \ref{fig:areaesigmapc} with the solid line representing the fit to the power law model \(\sigma_{pc}(p_0) \sim L^{-\gamma_0(p_0)}\). Here, one can observe that at the critical point of percolation, the conductivity presents a universal behavior with the scaling exponent assuming the value \(\gamma_0(p_0) = \gamma_0 = 0.754 \pm 0.004\) independent of \(p_0\).

For completeness, Table \ref{table:3} brings a comparison between the exponent \(\gamma_0\) and the value of the difference \(D_f-D_w\) for the values of \(p_0\) shown in the Table \ref{table:2}.

\begin{table}[h]
\centering
\begin{tabular}{|p{0.9cm}|p{3.0cm}|p{3.0cm}|} 
 \hline
 \(p_0\) & \(\gamma_0\) & \(D_f-D_w\) \\ [0ex] 
 \hline
0.2   &  \(-0.755 \pm 0.002\)  & \(-1.10 \pm 0.11\) \\
0.4   &  \(-0.753 \pm 0.003\)  & \(-1.10 \pm 0.10\) \\
0.6   &  \(-0.755 \pm 0.003\)  & \(-0.93 \pm 0.11\) \\
0.8   &  \(-0.754 \pm 0.002\)  & \(-0.75 \pm 0.07\) \\
1.0   &  \(-0.756 \pm 0.002\)  & \(-0.76 \pm 0.07\) \\ [0ex] 
 \hline
\end{tabular}
\caption{Comparison between the exponent \(\gamma_0\) and the values of the difference \(D_f-D_w\) for different values of $p_0$. The values are provided with their respective uncertainties.}
\label{table:3}
\end{table}

It is very interesting that, according to the obtained results, the Eq.\ (\ref{eq:Rscaling}) which is deduced from the Einstein relation, is valid only for \(p_0\) close to $1$ exactly where the \(D_f\) values are close to the values obtained for the case of classical percolation (see Table \ref{table:2}). 
At this point, it is important to mention that there are no works in the literature where the transport properties are studied in the way it was done in this work, so it is difficult to compare with results obtained by other authors.

\subsection{Fractal properties of the interface} \label{sect:Interface}

Finally, to highlight the complex interplay between the orientational anisotropy of the depositing dimer and the structure formation, we study the interfacial properties.

The interface profile is characterized by the curve that separates the multilayered structure from the surrounding environment at a given instant of the growth process.

\begin{figure}[t]
\begin{tabular}{c}
\includegraphics[width=.8\columnwidth]{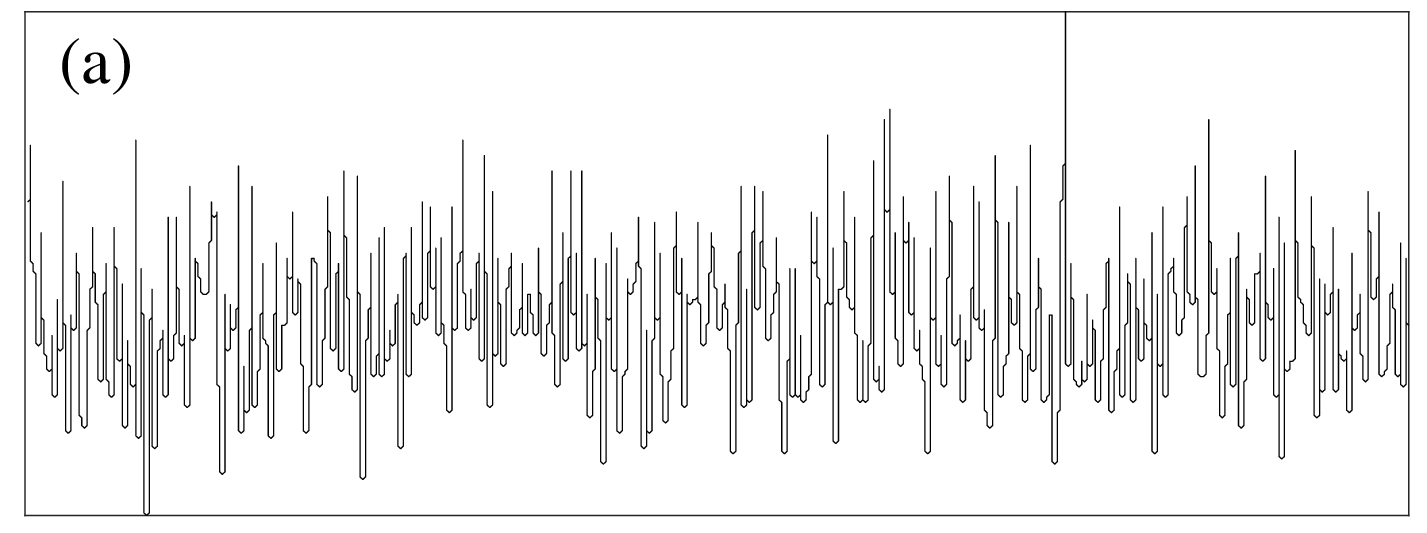} \\
\includegraphics[width=.8\columnwidth]{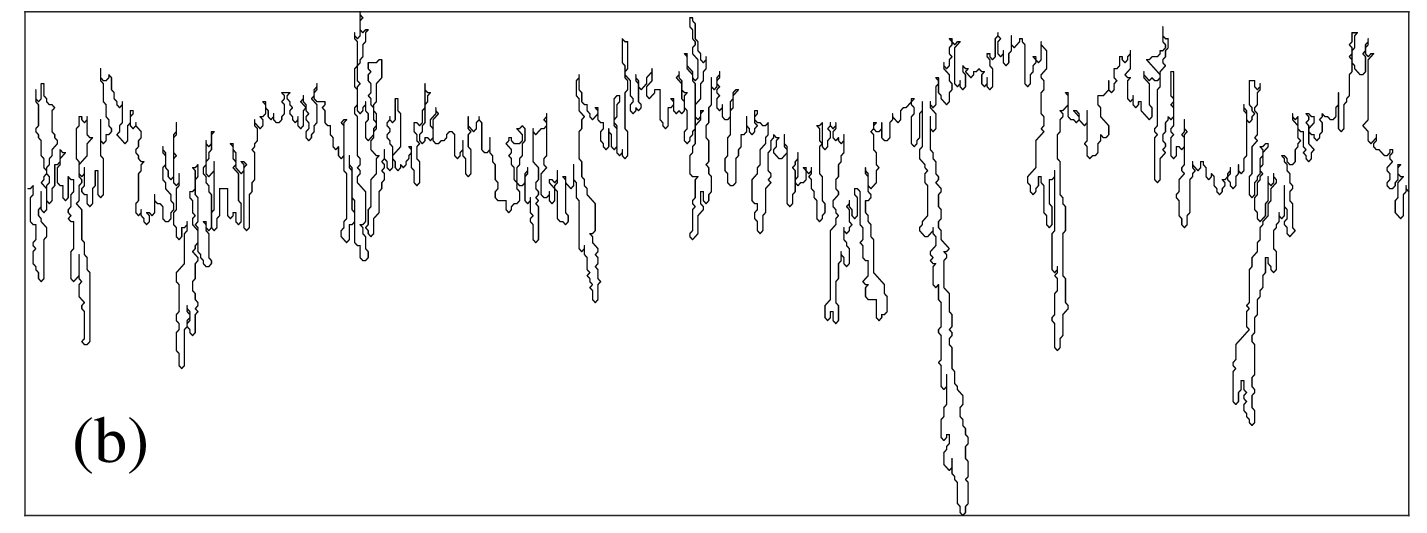} \\
\includegraphics[width=.8\columnwidth]{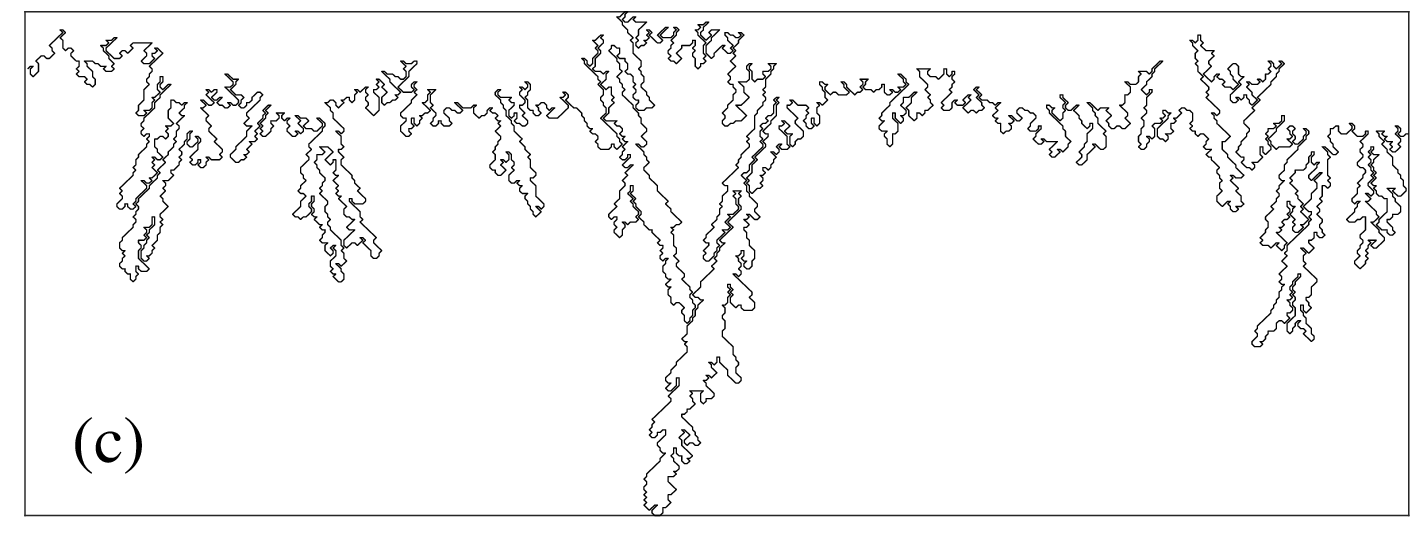}
\end{tabular}  
\caption{Typical interfaces or active regions for \(p_0 = 0\) (a), \(p_0=0.3\) (b), and \(p_0=1\) (c) using $H_{max}=L=512$, showing clearly the impact of orientational anisotropy of dimers on determining the interface of the growing structure.}
\label{fig:bounda}
\end{figure}

In the standard model of ballistic deposition of monomers by Meakin et.\ al.\/, the interface is defined as the bijective curve formed by the maximum heights of all columns. In contrast, here we define the interface as the non-bijective curve that traces the path of an ant walking on the stack of deposited dimers. Note that non-bijectivity refers to the fact that for a given horizontal position, there can generally be more than one possible values for the height corresponding to a specific column. This consideration provides us to uncover a more detailed understanding of the inherent characteristics of the interface. By varying the value of $p_0$, we examine the interface structure at a height of the formation $H_{max}=L$. Figure \ref{fig:bounda} displays typical examples of interface profiles for three different values of $p_0$. It is prominent that the orientational anisotropy of dimers has a great impact on the structure of the interface.

\begin{figure} [t] \centering  
\subfigure{\includegraphics[width=0.49\columnwidth]{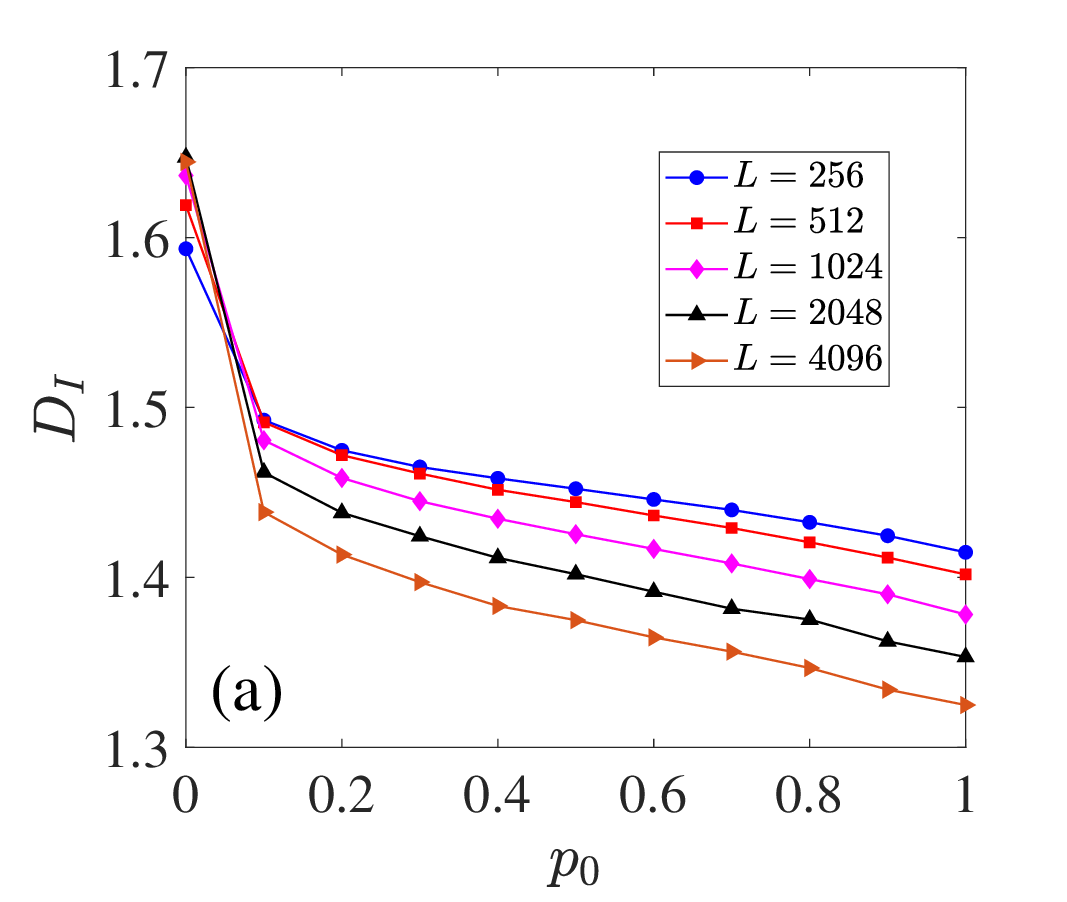}} 
\subfigure{\includegraphics[width=0.49\columnwidth]{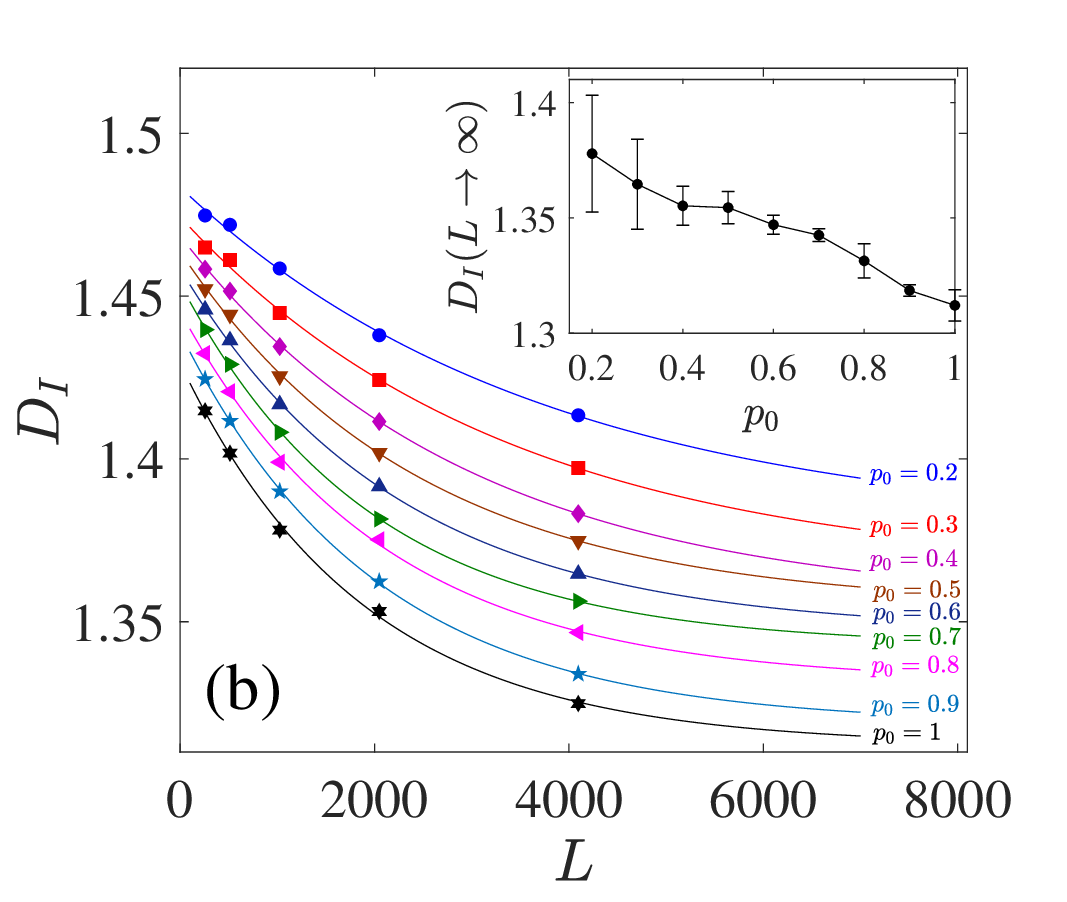}} 
    \caption{(a) The fractal dimension \(D_I\) of  the interface as function of the probability \(p_0\). (b) Asymptotic analysis of \(D_I\). The inset presents the behavior of the parameter \(c=D_I(L \to \infty)\) in Eq.\ \ref{eq:expmodel}, giving the value of the fractal dimension of the interface in the thermodynamic limit as a function of \(p_0\). In all cases, the error bar (variance) is much smaller than the size of the symbols. See text for detail.}
    \label{fig:fractalbound}
\end{figure}

For a quantitative description, we calculate the fractal dimension $D_I(L)$ of the interface for various values of $p_0$ on a lattice of size $L$. In Fig.\ \ref{fig:fractalbound}(a), we show the dependence of the fractal dimension $D_I(L)$ on the parameter $p_0$. For the determination of the fractal dimension, the one-dimensional Box-counting method was applied. It consists of measuring the perimeter of the curve with ``rulers" of size \(\epsilon\) and writing down the number \(N\) of possible rulers of this size that completely cover the curve. A log-log plot of \(N\) against the inverse of \(\epsilon\) then yields a straight line with slope $D_I(L)$ (not shown).  
For a reliable estimate, the data are averaged over $10^6$ independent replicas. It is also observed that for sufficiently large values of \(p_0\), the fractal dimension $D_I(L)$ decreases linearly with increasing $p_0$.

Asymptotic analysis \(D_I(L \to \infty)\) based on the same functional form adopted in subsection \ref{sect:Fractal} [Eq.\ (\ref{eq:expmodel})] is shown in the Fig.\ \ref{fig:fractalbound} (b). The inset exhibits the variation of \(D_I(L \to \infty)\) for \(p_0 \geq 0.2\).

As a closing remark to this section, we consider a few comments to be pertinent: (i) {We notice that the obtained values of $D_I(L \to \infty)$, exhibited in the inset of Fig.\ \ref{fig:fractalbound}(b), in a very wide interval of \(p_0\) are in agreement with the fractal dimension of the two-dimensional self-avoiding walks within typical uncertainties of less than 1\%. (ii) An equally intriguing result is that in the opposite direction, i.e. in the low-\(p_0\) region, the Fig.\ \ref{fig:fractalbound}(a) indicates that $D_I(L \to \infty)$ for the interface, in the limit \(p_0=0\), converges close to the value found for the backbone of the ordinary percolation problem (\(d_B = 1.62 \pm 0.02\)) \cite{Herrmann_Stanley_1984} at the percolation threshold. (iii) These findings (i) and (ii) concerning \(D_I\) in two different domains of \(p_0\) appear to be related, because the backbone of the percolation cluster consists of all the sites visited by all possible self-avoiding walks from the injection site(s) to the exit site(s)~\cite{Feder1988}.

\section{Conclusion} 
\label{sec:Canclu}
We numerically investigated the morphology of a growing multilayer two-dimensional structure built on a one-dimensional substrate that represents a simple not yet studied model of disordered and amorphous matters involving ballistic deposition of dimers with two possible orientations, horizontal and vertical, selected at random with probability $p_0$ and $(1-p_0)$, respectively. The most interesting characteristic of the multilayer packing structures of dimers studied here is the dendritic morphology as depicted in Figs.\ \ref{fig:varp0}, \ref{fig:culsters} and \ref{fig:Hcincrease}, with a labyrinthine pore structure that is dominated by the contribution of the complex geometry of both the internal and the boundary parts of the system. In this work, a great deal of effort is dedicated to deeply analyzing the physical properties of the percolation cluster, its critical behavior, nontrivial scaling laws, critical exponents in the thermodynamic limit (Sections \ref{sect:Percolation} and \ref{sect:Fractal}, Fig. \ref{fig:scalingHcandSmax}, Table \ref{table:1}) and the corresponding emergent electrical conductivity (Sec.\ \ref{sect:cunct}, Figs.\ \ref{fig:scalingsigma} and \ref{fig:areaesigmapc}, Tables \ref{table:2} and \ref{table:3}). The fractal aspects of the bulk and perimeter (Sec.\ \ref{sect:Interface}, Figs.\ \ref{fig:bounda} and \ref{fig:fractalbound}) of the structure have also been examined. 

As a future problem, one may consider a modified version of the deposition process by introducing defects into the system, general $k$-mers, and also, bringing up the diffusional and temperature effects that might influence the growth process itself, leading to describe a more realistic dynamics for a range of surface growth phenomena. Moreover, the present study can help researchers from other areas to quantify and control different levels of environmental concern linked to the accumulation of pollutants in trees, as well as in other natural and man-made structures, in places where contamination can reach different levels.

\section*{Acknowledgement}
G Palacios thanks a fellowship from Conselho Nacional de Desenvolvimento Científico e Tecnológico (CNPq) Process: 381191/2022-2. SK is supported by Research Grant No. `ORLA$\_$BIRD2020$\_$01' from the University of Padova. L A P Santos acknowledges CNPq for the grant 305017/2021-7. M A F Gomes acknowledges the financial support from the Brazilian Agency CAPES PROEX 23038.003069/2022-87, no. 0041/2022.

%\bibliography{ref}


\end{document}